# Game-Theoretic Pricing and Selection with Fading Channels

Yuqing Ni*, Alex S. Leong†, Daniel E. Quevedo†, and Ling Shi*

*Abstract*— We consider pricing and selection with fading channels in a Stackelberg game framework. A channel server decides the channel prices and a client chooses which channel to use based on the remote estimation quality. We prove the existence of an optimal deterministic and Markovian policy for the client, and show that the optimal policies of both the server and the client have threshold structures when the time horizon is finite. Value iteration algorithm is applied to obtain the optimal solutions for both the server and client, and numerical simulations and examples are given to demonstrate the developed result.

## I. INTRODUCTION

With the increasing demands of wireless communication service, Networked Control Systems (NCSs) have been replacing the traditional wired network control systems. NCSs utilize wireless channels to communicate between sensors and estimators, and controllers and actuators [1]. It is well known that compared to wired systems, NCSs have a lower cost in installment, diagnosis, debugging and maintenance, and have diverse applications, including manufacturing industries, public transportation, aerospace vehicles and even military battles [2].

Sensors with sensing, computing and wireless communication capabilities measure ambient conditions and transmit extracted and useful signals to remote estimators via wireless channels [3]. However, transmission may suffer from packet dropouts, delay, quantization and channel fading, which affect the performance and stability of real time systems. Transmission errors in physical network links and buffer overflows due to congestion are the typical reasons for packet dropouts [4]. Sinopoli et al. [5] show that if the average packet arrival probability does not reach a certain threshold, the system may diverge and will not be under control any more.

Nowadays, timely and accurate information is of great value. As the transmission service provider, the server should determine the communication channel prices. The wireless channel pricing for the server and channel selections for the client (which here means the sensor) are important. Game theory is introduced as a powerful tool to study the rational players' complex interactions. A player is rational if his decisions are consistent with his objectives [6]. Generally, a Nash equilibrium is an optimal solution such that no player can benefit more by changing his own strategy [7]. While a Stackelberg game is a strategic game where players make decisions sequentially. Saad et al. [8] apply game theory to communication networks and consider cooperative scenarios, referred to as coalitional game theory. Li et al. [9] consider the remote state estimation of cyber-physical systems (CPSs) under denial-of-service (DoS) attacks in a Markov game framework. They analyze the Nash equilibrium of a zero-sum game. The issue of network pricing and bandwidth allocation is addressed in [10] based on the Nash bargaining framework. They show that the network revenue is maximized when the pricing scheme depends on users' budgets and bandwidth demands. Pricing and allocation related problems are also studied in [11] and [12]. Tushar et al. [13] study the pricing and charging problem of a smart grid and plug-in electric vehicle groups in a noncooperative Stackelberg game framework. They prove the existence of a generalized Stackelberg equilibrium (GSE) using convex optimization and provide an algorithm which leads to the GSE state of the game. However, scenarios in NCSs feature a dynamic process with infinite time horizon. It is not a static optimization problem.

In this work, we investigate a remote estimation problem with wireless communication channels. A Stackelberg game framework is modeled to interpret interactions between the sensor client and the channel server. The channel server aims to maximize its revenue while the sensor client aims to minimize the linear combination of remote estimation error covariance and the cost in a sequential order. We analyze the optimal policies for both players when they reach the equilibrium.

There are related applications in real life. Take Google Drive for an example, which provides 15 GB storage free of charge [14]. Premium plans are available for users with more storage space demands. For example, $1.99 / month with 100 GB, $9.99 / month with 1 TB and $99.99 / month with 10 TB are all upgraded alternatives. Obviously, network users are more and more in need of cloud storage, online sharing and backup. Some users are willing to pay more to enjoy the privilege. This scenario can be modeled as a Stackelberg game. As the resource holder, Google Drive is at a leading position, who sets the rules first. To profit most, Google Drive should ponder over its service and corresponding charge, taking the general user market into consideration in advance. While users decide whether to pay more based on their own needs. Both players intend to maximize their benefits.

The remainder of the paper is organized as follows. Section II introduces the system model and the main problem. Section III proves the existence and monotonicity of the

*The work by Y. Ni and L. Shi is supported by a Hong Kong RGC General Research Fund 16222716.

*: Department of Electronic and Computer Engineering, Hong Kong University of Science and Technology, Clear Water Bay, Kowloon, Hong Kong (email: yniac@ust.hk, eesling@ust.hk).

†: Department of Electrical Engineering (EIM-E), Paderborn University, Paderborn, Germany (email: alex.leong@upb.de, dquevedo@ieee.org)

optimal deterministic and Markovian policy in the finite time horizon case. Section IV provides simulations and intuitive interpretations. Section V draws conclusions.

*Notation*: For a matrix $X$, we use $X^T$ and $\text{Tr}\{X\}$ to denote its transpose and trace. When $X$ is a positive semi-definite matrix, it is written as $X \geq 0$. For two symmetric matrices $X$ and $Y$, $X \geq Y$ means $X - Y \geq 0$. $\mathbb{R}$ is the set of real numbers. $\mathbb{E}[\cdot]$ is the expectation of a random matrix and $\mathbb{E}[\cdot|\cdot]$ is its conditional expectation. The notation $\mathbb{P}[\cdot]$ refers to probability. For functions $f, f_1, f_2$, $f_1 \circ f_2$ is defined as $f_1 \circ f_2(X) \triangleq f_1(f_2(X))$ and $f^k$ is defined as $f^k(X) \triangleq \underbrace{f \circ f \cdots f}_{k \text{ times}}(X)$ with $f^0(X) = X$.

## II. PROBLEM SETUP

### A. System Model

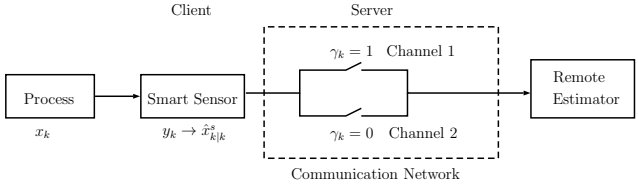

Fig. 1. System block diagram

Consider the system in Fig. 1. The discrete linear time-invariant (LTI) process is as follows:

$$x_{k+1} = Ax_k + w_k, \quad (1)$$
$$y_k = Cx_k + v_k, \quad (2)$$

where $x_k \in \mathbb{R}^n$ is the process state vector, $w_k \in \mathbb{R}^n$ is the process noise which is i.i.d. zero-mean white Gaussian with covariance $Q \geq 0$. The measurement collected by the sensor is $y_k \in \mathbb{R}^m$. The measurement noise $v_k \in \mathbb{R}^m$ is i.i.d. zero-mean white Gaussian with covariance $R > 0$. The initial state $x_0$ is a zero-mean Gaussian random variable with covariance $\Pi_0 \geq 0$, which is uncorrelated with $w_k$ and $v_k$. The pair $(A,C)$ is assumed to be observable and $(A, \sqrt{Q})$ is controllable.

The smart sensor in Fig. 1 is capable of running a local Kalman filter. Its minimum mean-squared error state estimate $\hat{x}^s_{k|k}$ and the corresponding error covariance $P^s_{k|k}$ are

$$\hat{x}^s_{k|k} = \mathbb{E}[x_k \mid y_1, \ldots, y_k], \quad (3)$$
$$P^s_{k|k} = \mathbb{E}\left[\left(x_k - \hat{x}^s_{k|k}\right)\left(x_k - \hat{x}^s_{k|k}\right)^T \mid y_1, \ldots, y_k\right], \quad (4)$$

which are computed via a Kalman filter as follows:

$$\hat{x}^s_{k|k-1} = A\hat{x}^s_{k-1|k-1}, \quad (5)$$
$$P^s_{k|k-1} = AP^s_{k-1|k-1}A^T + Q, \quad (6)$$
$$K_k = P^s_{k|k-1}C^T\left[CP^s_{k|k-1}C^T + R\right]^{-1}, \quad (7)$$
$$\hat{x}^s_{k|k} = \hat{x}^s_{k|k-1} + K_k\left(y_k - C\hat{x}^s_{k|k-1}\right), \quad (8)$$
$$P^s_{k|k} = P^s_{k|k-1} - K_k CP^s_{k|k-1}. \quad (9)$$

Anderson and Moore [15] show that the estimation error covariance of the Kalman filter converges to a unique value $\overline{P}$ no matter what the initial value is. Define the Lyapunov operator $h$ as $h(X) \triangleq AXA^T + Q$ and Riccati operator $\tilde{g}$ as $\tilde{g}(X) \triangleq X - XC^T\left[CXC^T + R\right]^{-1}CX$. We assume that the error covariance at the smart sensor has already reached steady state and let

$$P^s_{k|k} = \overline{P}, \ k \geq 0, \quad (10)$$

where $\overline{P}$ is the unique positive semi-definite solution to $\tilde{g} \circ h(X) = X$ as [15] shows.

The error covariance $\overline{P}$ satisfies:

$$h^{k_1}(\overline{P}) \geq h^{k_2}(\overline{P}) \quad (11)$$

for any $k_1 \geq k_2 \geq 0$ (see [16]). Furthermore,

$$\text{Tr}\left\{h^{k_1}(\overline{P})\right\} \geq \text{Tr}\left\{h^{k_2}(\overline{P})\right\}. \quad (12)$$

This well ordering of the estimation error covariance is helpful to the further analysis.

Fig. 1 portrays a two-player scenario. The process equipped with the smart sensor is the "client" and the communication network belongs to the "server". At each time $k$, the server has the priority to decide the price of the high quality channel 1, i.e., $W_H$ or $W_L$. The price of the low quality channel 2 is fixed, always $W_0$. Without loss of generality, we assume $W_H > W_L > W_0$. It is then the client's turn to decide to use channel 1 or channel 2. If using channel 1, the probability that $\hat{x}^s_{k|k}$ arrives error-free at the remote estimator is $\lambda_1$. While using channel 2, the error-free arrival probability is $\lambda_2$, where $1 > \lambda_1 > \lambda_2 > 0$.

The server sets the price of channel 1. Define the price $W_k$ of channel 1 at time $k$ as

$$W_k = \begin{cases} W_H, & \text{if the server sets high price,} \\ W_L, & \text{if the server sets low price.} \end{cases} \quad (13)$$

The client decides whether to use channel 1 or not. Thus the transmission of $\hat{x}^s_{k|k}$ in the communication network can be characterized by a binary variable $\gamma_k$:

$$\gamma_k = \begin{cases} 1, & \text{if the client uses channel 1,} \\ 0, & \text{if the client uses channel 2.} \end{cases} \quad (14)$$

Furthermore, the arrival of packets at the remote estimator can be characterized by a binary random sequence $\{\delta_k\}$:

$$\delta_k = \begin{cases} 1, & \text{if } \hat{x}^s_{k|k} \text{ arrives error-free at time } k, \\ 0, & \text{otherwise.} \end{cases} \quad (15)$$

Based on the error-free arrival probability of the two channels, we have

$$\mathbb{P}[\delta_k = 1] = \gamma_k\lambda_1 + (1 - \gamma_k)\lambda_2. \quad (16)$$

## B. Remote Estimator

Denote $\hat{x}_k$ and $P_k$ as the state estimate and error covariance at the remote estimator. If the packet $\hat{x}^s_{k|k}$ arrives error-free at time $k$, the estimator synchronizes $\hat{x}_k$ with $\hat{x}^s_{k|k}$ from the smart sensor; otherwise, it just uses the time update value based on the system model (1). The recursion of $\hat{x}_k$ is

$$\hat{x}_k = \begin{cases} \hat{x}^s_{k|k}, & \text{if } \delta_k = 1, \\ A\hat{x}_{k-1}, & \text{if } \delta_k = 0. \end{cases} \quad (17)$$

Correspondingly, the recursion of $P_k$ is

$$P_k = \begin{cases} \overline{P}, & \text{if } \delta_k = 1, \\ h(P_{k-1}), & \text{if } \delta_k = 0. \end{cases} \quad (18)$$

To simplify the problem, we assume that the initial state $P_0 = \overline{P}$. At each time $k$, $P_k$ takes a value from the set $\{\overline{P}, h(\overline{P}), h^2(\overline{P}), \ldots\}$. Note that the server is the owner of the communication network and it has a good knowledge of the channel transmission process. The server always knows exactly whether $\hat{x}^s_{k|k}$ arrives at the remote estimator successfully. We assume that the remote estimator sends an Acknowledgment (ACK) to the client. The ACK is only a 1-bit signal to inform the client that whether the data packet arrives successfully or not. Thus this is a perfect-information case for both the server and the client.

## III. FINITE TIME HORIZON GAME FRAMEWORK

Consider the following optimization problem in a Markov game framework. In the finite time horizon ($k = 1, 2, \ldots, N$) case, for the client, its objective function $J_C$ is a linear combination of the trace of expected estimation error covariance and the cost when using different channels:

$$\max_{\{\gamma_k\}} J_C \triangleq -\sum_{k=1}^{N}\left[\zeta \text{Tr}\{\mathbb{E}[P_k]\} + (1-\zeta)[\gamma_k W_k + (1-\gamma_k)W_0]\right], \quad (19)$$

for some weight parameter $\zeta \in (0,1)$. A larger $\zeta$ attaches more importance to the error covariance and a smaller $\zeta$ attaches more importance to the channel costs. For every $\{W_k\}$ given by the server, the client needs to determine the optimal action $\{\gamma_k\}$ to maximize its $J_C$. While for the server, its objective function $J_S$ is the total revenue:

$$\max_{\{W_k\}} J_S \triangleq \sum_{k=1}^{N}\left[\gamma_k W_k + (1-\gamma_k)W_0\right]. \quad (20)$$

The server is at the leading position in this Stackelberg game. At each time $k$, it sets the price $W_k$ first and then the follower, the client, decides $\gamma_k$ sequentially. We define $\mathscr{I}^S_k$ as the information set available to the server up to time $k$, i.e., $\mathscr{I}^S_k = \{P_0, P_1, \ldots, P_{k-1}\}$ and $\mathscr{I}^C_k = \{P_0, P_1, \ldots, P_{k-1}\} \cup \{W_1, W_2, \ldots, W_k\}$ for the client. Assume both players are rational, which means that they take the optimal actions based on their information sets respectively.

## A. Client's Optimal Policy

In this section, we prove that for each fixed $\{W_k\}$, the client's optimal policy is deterministic and Markovian, and has a threshold on $P_{k-1}$.

For each given $\{W_k\}$, the remote estimator's state set observed by the client is $\mathbb{S} = \{\text{Tr}\{\overline{P}\}, \text{Tr}\{h(\overline{P})\}, \text{Tr}\{h^2(\overline{P})\}, \ldots, \text{Tr}\{h^N(\overline{P})\}\}$ due to the finite time horizon $N$, and the client's action set is $\mathbb{A} = \{0, 1\}$. We formulate this optimization problem as a Markov decision process (MDP) problem. The stochastic kernel of the state at the remote estimator is given by:

$$p_k(\text{Tr}\{P_k\} \mid \text{Tr}\{P_{k-1}\}, \gamma_k) \triangleq$$
$$\begin{cases} \gamma_k \lambda_1 + (1-\gamma_k)\lambda_2, & \text{if } \text{Tr}\{P_k\} = \text{Tr}\{\overline{P}\}, \\ 1 - \gamma_k \lambda_1 - (1-\gamma_k)\lambda_2, & \text{if } \text{Tr}\{P_k\} = \text{Tr}\{h(P_{k-1})\}. \end{cases} \quad (21)$$

It is obvious that the stochastic kernel $p_k(\cdot \mid \cdot, \cdot)$ is stationary, which means it is only dependent on the state and the action. We can substitute $p(\cdot \mid \cdot, \cdot)$ for it.

The one-stage reward function for the client at time $k = 1, 2, \ldots, N$ can be written as:

$$r_k(\text{Tr}\{P_{k-1}\}, \gamma_k) = -\Big[\zeta \text{Tr}\{\mathbb{E}[P_k]\} \\ + (1-\zeta)[\gamma_k W_k + (1-\gamma_k)W_0]\Big]. \quad (22)$$

Then the client's optimization problem can be transformed equivalently to an MDP problem as follows:

$$\max_{\{\gamma_k\}} \mathbb{E}\left[\sum_{k=1}^{N} r_k(\text{Tr}\{P_{k-1}\}, \gamma_k)\right] \quad (23)$$

For simplicity, we define the neighbouring element after and before $s$ in the state set $\mathbb{S} = \{\text{Tr}\{\overline{P}\}, \text{Tr}\{h(\overline{P})\}, \text{Tr}\{h^2(\overline{P})\}, \ldots, \text{Tr}\{h^N(\overline{P})\}\}$ as $s^+, s^-$ respectively, and define $\text{Tr}\{\overline{P}\}$ as $s_0$.

Define the optimality equation (also referred to as Bellman equation or functional equation) as

$$V_k(s) = \sup_{a \in \mathbb{A}} \Big\{ r_k(s,a) + [p(s_0 \mid s,a)V_{k+1}(s_0) \\ + p(s^+ \mid s,a)V_{k+1}(s^+)] \Big\} \quad (24)$$

for $k = 1, 2, \ldots, N$ where $V_k(\cdot)$ is a real-valued function and $V_{N+1}(s) = 0$. The optimality equations are fundamental tools to study MDP problems [17]. If the expected total reward of a policy from period $k$ onward satisfies these optimality equations for $k = 1, 2, \ldots, N$, then it is optimal.

**Theorem 1** *Assume that the state set $\mathbb{S}$ is finite, and that the action set $\mathbb{A}$ is finite for each state $s \in \mathbb{S}$. Then there exists a deterministic Markovian policy which is optimal.*

*Proof:* Let $V^*_k$ be a solution of the optimality equation (24). For each $k$ and $s \in \mathbb{S}$, if there exists an $a' \in \mathbb{A}$ such that

$$r_k(s,a') + [p(s_0|s,a')V^*_{k+1}(s_0) + p(s^+|s,a')V^*_{k+1}(s^+)] \\ = \sup_{a \in \mathbb{A}} \{r_k(s,a) + [p(s_0|s,a)V^*_{k+1}(s_0) + p(s^+|s,a)V^*_{k+1}(s^+)]\}, \quad (25)$$

then there exists an optimal policy which is deterministic and Markovian. The proof is given by Theorem 4.4.2 in [17]. Here clearly such an $a'$ always exists if the action set $\mathbb{A}$ is finite. ∎

The finite state set $\mathbb{S}$ and action set $\mathbb{A}$ in our problem determine the existence of a deterministic optimal policy. In addition, it is Markovian.

Next we prove that the client's optimal policy under each given $\{W_k\}$ has a threshold structure based on the state. To be more specific, we prove that there exists a positive $\varepsilon_k$ such that when the current state $\text{Tr}\{P_{k-1}\} \geq \varepsilon_k$, the optimal policy for the client is to use channel 1. This result has an intuitive meaning. If the remote estimation error covariance is extremely large, the client should choose the high quality channel 1 to ensure a higher packet arrival probability. Otherwise, the client will choose channel 2 to obtain a lower cost. Thus there is a trade-off between estimation quality and transmission cost. To prove this, some preliminary knowledge is needed.

**Definition 1** *For the well ordered sets X and Y and a real-valued function $g(x,y)$ on $X \times Y$, $g$ is superadditive if for $x^+ \geq x^-$ in X and $y^+ \geq y^-$ in Y,*

$$g(x^+, y^+) + g(x^-, y^-) \geq g(x^+, y^-) + g(x^-, y^+). \quad (26)$$

**Lemma 1** *Suppose $g$ is a superadditive function on $X \times Y$ and for each $x \in X$, $\max_{y \in Y} g(x,y)$ exists. Then*

$$f(x) = \max\left\{y^* \in \arg\max_{y \in Y} g(x,y)\right\} \quad (27)$$

*is monotone nondecreasing in x.*

*Proof:* Choose $y \leq f(x^-)$. Due to the definition of superadditivity and $f(x)$,

$$g(x^+, f(x^-)) + g(x^-, y) \geq g(x^+, y) + g(x^-, f(x^-)), \quad (28)$$

and

$$g(x^-, f(x^-)) \geq g(x^-, y). \quad (29)$$

Rewriting (28) and combining it with (29) yields

$$g(x^+, f(x^-)) \geq g(x^+, y) + [g(x^-, f(x^-)) - g(x^-, y)] \geq g(x^+, y) \quad (30)$$

for all $y \leq f(x^-)$. Due to the definition of $f(x)$,

$$g(x^+, f(x^+)) \geq g(x^+, f(x^-)). \quad (31)$$

Consequently, $f(x^+) \geq f(x^-)$. The proof is done. ∎

We demonstrate the optimality of monotone policy by showing that the optimal value function $V_k^*(s)$ for all $k$ is nonincreasing in $s \in \mathbb{S}$ and then showing that

$$\Phi_k(s,a) \triangleq r_k(s,a) + [p(s_0 \mid s,a)V_{k+1}^*(s_0) + p(s^+ \mid s,a)V_{k+1}^*(s^+)] \quad (32)$$

is superadditive.

**Theorem 2** *The optimal value function $V_k^*(s)$ is nonincreasing in $s \in \mathbb{S}$ for $k = 1, 2, \ldots, N$.*

*Proof:* Notice that in the finite time horizon problem, we have $V_{N+1}^*(s) = 0$. It is trivial to see that $V_{N+1}^*(s)$ is nonincreasing in $s$. We prove the monotonicity in a backward induction way.

Assume that $V_t^*(s)$ is nonincreasing for $t = k+1, \ldots, N+1$. By Theorem 1, there exists an optimal $a_s^* \in \mathbb{A}$ which obtains the maximum of $V_k(s)$:

$$V_k^*(s) = r_k(s, a_s^*) + [p(s_0 \mid s, a_s^*)V_{k+1}^*(s_0) + p(s^+ \mid s, a_s^*)V_{k+1}^*(s^+)]. \quad (33)$$

Let $s' \geq s$. Due to the well ordered state set $\mathbb{S}$, it is easy to verify that $r_k(s,a)$ is nonincreasing in $s$ for all $a \in \mathbb{A}$ and $k = 1, \ldots, N$ and the stochastic kernel $p(s^+ \mid s,a)$ and $p(s_0 \mid s,a)$ are only dependent on $a$ for all $s \in \mathbb{S}$.

$$V_k^*(s) = \max_{a \in \mathbb{A}} \left\{ r_k(s,a) + [p(s_0|s,a)V_{k+1}^*(s_0) + p(s^+|s,a)V_{k+1}^*(s^+)] \right\}$$
$$\geq r_k(s, a_{s'}^*) + [p(s_0|s, a_{s'}^*)V_{k+1}^*(s_0) + p(s^+|s, a_{s'}^*)V_{k+1}^*(s^+)]$$
$$\geq r_k(s', a_{s'}^*) + [p(s_0|s', a_{s'}^*)V_{k+1}^*(s_0) + p(s'^+|s', a_{s'}^*)V_{k+1}^*(s'^+)]$$
$$= V_k^*(s'). \quad (34)$$

Thus $V_k^*(s)$ is nonincreasing for $k = 1, 2, \ldots, N$. ∎

**Theorem 3** *$\Phi_k(s,a)$ is a superadditive function on $\mathbb{S} \times \mathbb{A}$.*

*Proof:* First we prove that $r_k(s,a)$ is superadditive.

$$[r_k(s,1) + r_k(s^-, 0)] - [r_k(s,0) + r_k(s^-, 1)]$$
$$= -\zeta[\lambda_1 s_0 + (1-\lambda_1)s^+] - (1-\zeta)W_k$$
$$\quad -\zeta[\lambda_2 s_0 + (1-\lambda_2)s] - (1-\zeta)W_0$$
$$\quad +\zeta[\lambda_2 s_0 + (1-\lambda_2)s^+] + (1-\zeta)W_0 \quad (35)$$
$$\quad +\zeta[\lambda_1 s_0 + (1-\lambda_1)s] + (1-\zeta)W_k$$
$$= \zeta(\lambda_1 - \lambda_2)(s^+ - s)$$
$$\geq 0,$$

which satisfies the definition of superadditive.

Since the sum of two superadditive functions is also superadditive, we need to prove that the rest of $\Phi_k(s,a)$ is superadditive. Together with the nonincreasing property of $V_k^*(s)$ from Theorem 2, we have

$$(p(s_0 \mid s, 1)V_{k+1}^*(s_0) + p(s^+ \mid s, 1)V_{k+1}^*(s^+)$$
$$\quad + p(s_0 \mid s^-, 0)V_{k+1}^*(s_0) + p(s \mid s^-, 0)V_{k+1}^*(s))$$
$$- (p(s_0 \mid s, 0)V_{k+1}^*(s_0) + p(s^+ \mid s, 0)V_{k+1}^*(s^+)$$
$$\quad + p(s_0 \mid s^-, 1)V_{k+1}^*(s_0) + p(s \mid s^-, 1)V_{k+1}^*(s))$$
$$= (\lambda_2 - \lambda_1)(V_{k+1}^*(s^+) - V_{k+1}^*(s))$$
$$\geq 0. \quad (36)$$

The proof is done. ∎

Thus we can conclude from Theorem 3 and Lemma 1 that the optimal policy for the client is nondecreasing in the state at each time epoch. In other words, there is a threshold based on the state when taking the optimal policy. For the finite time horizon case, it is feasible to obtain this deterministic Markovian optimal policy by value iteration and obtain its threshold.

## B. Server's Optimal Policy

The server is the leading player in this Stackelberg game. In this section, we show that the server's optimal policy is also monotone. The larger the estimation error covariance is, the higher the price can be set.

In Theorem 4 below, we consider an extension to the current problem formulation where the server can set any price $W_k \geq W_0$ for channel 1.

**Theorem 4** *The optimal price $W_k^*$ is nondecreasing in $P_{k-1}$ for $k = 1, 2, \ldots, N$.*

*Proof:* Define the functions $U_k(\cdot, \cdot) : \mathbb{S} \times \mathbb{R} \to \mathbb{R}$ in a recursive way:

$$U_{N+1}(P_N) = 0, \quad (37)$$

$$U_k(P_{k-1}, W_k) = \max_{\gamma_k} \Big\{ -[\zeta \text{Tr}\{\mathbb{E}[P_k]\} + (1-\zeta)[\gamma_k W_k + (1-\gamma_k)W_0]] + [\gamma_k \lambda_1 + (1-\gamma_k)\lambda_2] U_{k+1}(\overline{P}, W_{k+1}) + [\gamma_k(1-\lambda_1) + (1-\gamma_k)(1-\lambda_2)] U_{k+1}(h(P_{k-1}), W_{k+1}) \Big\} \quad (38)$$

for $k = 1, 2, \ldots, N$. We prove it using induction. Consider first the time $N$. If $\gamma_N = 1$,

$$U_N(P_{N-1}, W_N) = -\Big[\zeta[\lambda_1 \text{Tr}\{\overline{P}\} + (1-\lambda_1)\text{Tr}\{h(P_{N-1})\}] + (1-\zeta)W_N\Big]. \quad (39)$$

If $\gamma_N = 0$,

$$U_N(P_{N-1}, W_N) = -\Big[\zeta[\lambda_2 \text{Tr}\{\overline{P}\} + (1-\lambda_2)\text{Tr}\{h(P_{N-1})\}] + (1-\zeta)W_0\Big]. \quad (40)$$

The difference of (39) and (40) is

$$-(1-\zeta)(W_N - W_0) - \zeta(\lambda_1 - \lambda_2)\left(\text{Tr}\{\overline{P}\} - \text{Tr}\{h(P_{N-1})\}\right). \quad (41)$$

The server wishes the client to use channel 1 and at the same time, to set the price as high as possible. This implies:

$$(1-\zeta)(W_N - W_0) \leq \zeta(\lambda_1 - \lambda_2)\left(\text{Tr}\{h(P_{N-1})\} - \text{Tr}\{\overline{P}\}\right). \quad (42)$$

Define the price threshold as

$$W_N^* = \frac{\zeta}{1-\zeta}(\lambda_1 - \lambda_2)\left(\text{Tr}\{h(P_{N-1})\} - \text{Tr}\{\overline{P}\}\right) + W_0, \quad (43)$$

which is the highest price that can be set to ensure the client to use channel 1. It can be seen that the larger $P_{N-1}$ is, the larger the price threshold $W_N^*$. As a consequence, the smaller $U_N(P_{N-1}, W_N^*)$ is.

Now, assume that $W_t^*$ is nondecreasing in $P_{t-1}$ and $U_t(P_{t-1}, W_t^*)$ is nonincreasing in $P_{t-1}$ for $t = k+1, \ldots, N$.

Then consider the time $k$. If $\gamma_k = 1$,

$$U_k(P_{k-1}, W_k) = -\Big[\zeta[\lambda_1 \text{Tr}\{\overline{P}\} + (1-\lambda_1)\text{Tr}\{h(P_{k-1})\}] + (1-\zeta)W_k\Big] + \lambda_1 U_{k+1}(\overline{P}, W_{k+1}^*) + (1-\lambda_1)U_{k+1}(h(P_{k-1}), W_{k+1}^*). \quad (44)$$

If $\gamma_k = 0$,

$$U_k(P_{k-1}, W_k) = -\Big[\zeta[\lambda_2 \text{Tr}\{\overline{P}\} + (1-\lambda_2)\text{Tr}\{h(P_{k-1})\}] + (1-\zeta)W_0\Big] + \lambda_2 U_{k+1}(\overline{P}, W_{k+1}^*) + (1-\lambda_2)U_{k+1}(h(P_{k-1}), W_{k+1}^*). \quad (45)$$

Similarly, the price threshold $W_k^*$ is

$$W_k^* = \frac{\zeta}{1-\zeta}(\lambda_1 - \lambda_2)\left(\text{Tr}\{h(P_{k-1})\} - \text{Tr}\{\overline{P}\}\right) + W_0 + \frac{1}{1-\zeta}(\lambda_1 - \lambda_2)\left(U_{k+1}(\overline{P}, W_{k+1}^*) - U_{k+1}(h(P_{k-1}), W_{k+1}^*)\right). \quad (46)$$

Then $W_k^*$ is nondecreasing in $P_{k-1}$ and $U_k(P_{k-1}, W_k^*)$ is nonincreasing in $P_{k-1}$ by the induction hypothesis. This monotonicity holds for all $k = 1, 2, \ldots, N$. ∎

For the original $W_H$ and $W_L$ case, the server sets the price

$$W_k = \begin{cases} W_H, & \text{if } W_H \leq W_k^*, \\ W_L, & \text{if } W_L \leq W_k^* < W_H, \\ W_H \text{ or } W_L, & \text{if } W_k^* < W_L. \end{cases} \quad (47)$$

Thus the server's optimal policy at each time is characterized in Fig. 2.

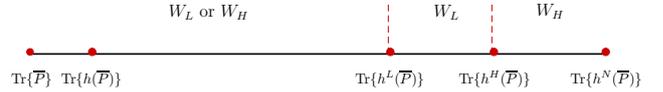

Fig. 2. Server's optimal policy

As the game leader, the server knows that at each given $\{W_k\}$, the client makes decisions based on the current state. And the server can obtain the policy thresholds $\text{Tr}\{h^L(\overline{P})\}$ and $\text{Tr}\{h^H(\overline{P})\}$ for $W_L$ or $W_H$ by calculation. When the current state $\text{Tr}\{P_{k-1}\} \geq \text{Tr}\{h^H(\overline{P})\}$, the server knows that the client chooses channel 1 even if the price is $W_H$. Thus setting $W_H$ maximizes the server's objective function. When $\text{Tr}\{h^L(\overline{P})\} \leq \text{Tr}\{P_{k-1}\} < \text{Tr}\{h^H(\overline{P})\}$, setting $W_L$ is the server's optimal policy. When $\text{Tr}\{P_{k-1}\} < \text{Tr}\{h^L(\overline{P})\}$, no matter what price the server sets, the state is still good and the client always chooses channel 2.

## C. Equilibrium Analysis

This is a Stackelberg game model, where the server has priority over the client. For both the server and the client, they make decisions based on the current state. In the first period, the server sets the price. This decision is irreversible. In the second period, the client chooses a channel after observing the price given by the server. By value iteration, the threshold can be calculated and both players can obtain their optimal policies to maximize their objectives. This is the equilibrium in this scenario.

## IV. SIMULATION

In this section, numerical examples are provided to illustrate the optimal policies for both players.

Consider the example with parameters of the unstable system where $A = \begin{bmatrix} 1.2 & 0 \\ 0 & 0.9 \end{bmatrix}, C = \begin{bmatrix} 1 & 0 \end{bmatrix}, Q = \begin{bmatrix} 0.3 & 0 \\ 0 & 0.3 \end{bmatrix}, R = 0.3$ and communication network channels where $\lambda_1 = 0.99, \lambda_2 = 0.2$. Set the weight parameter $\zeta = 0.5$. Assume $W_0 = 5$ and set $W_H = 100$, $W_L = 10$. Let the server's pricing strategy be $\{W_k = W_L\}$ and $\{W_k = W_H\}$ respectively. Now we apply the value iteration algorithm to the finite time horizon where $N = 20$. The client's optimal monotone policy of $\{W_L\}$ and $\{W_H\}$ at time step $k = 4$ is shown in Fig. 3. Whether at $W_H$ or $W_L$, the client's optimal policy is monotone in the state.

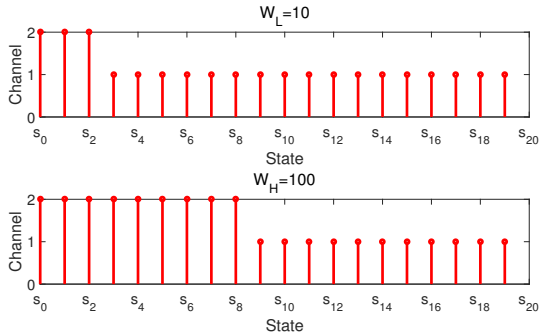

Fig. 3. Client's optimal policy $\gamma_k^*$ for different states under $W_L$ and $W_H$ at time $k = 4$

Consider the server's optimal pricing setting strategy. Assume that $W_0 = 5$, $W_L = 5.3$ and $W_H = 5.5$, and the finite time horizon is $N = 5$. The server's optimal policy is plotted in Fig. 4. It can be seen that the server's optimal policy is monotone in the state at each time. The worse the current state is, the higher the price can be set. This "Fishing in Troubled Water" action makes server benefit most from the misfortunes of the client. However, the price cannot be set too high in case the client does not choose channel 1. Thus this optimal policy is of significance to the server.

## V. CONCLUSIONS

We considered channel pricing and selection in a Stackelberg game framework. Using tools from Markov decision process, we proved the existence of an optimal deterministic and Markovian policy for the client and its monotone property in the finite time horizon case. Value iteration can be applied to calculate optimal policies both for the server and client. Future work includes considering the infinite time horizon case, and channel pricing and selection with multiple clients where the number of clients affects the transmission quality of channels. Also, imperfect information Stackelberg game can be explored, which induces a partially observable Markov decision process problem.

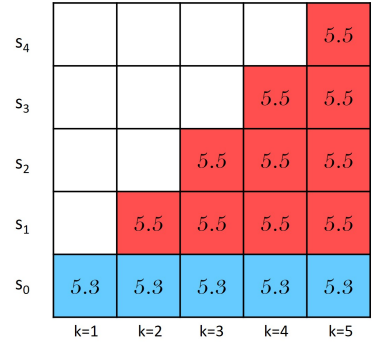

Fig. 4. Server's optimal policy $W_k^*$ for different states when $N = 5$

## REFERENCES

[1] R. A. Gupta and M.-Y. Chow, "Networked control system: Overview and research trends," *IEEE transactions on industrial electronics*, vol. 57, no. 7, pp. 2527–2535, 2010.
[2] R. M. Murray, *Control in an Information Rich World: Report of the Panel on Future Directions in control, Dynamics, and Systems*. SIAM, 2003.
[3] J. N. Al-Karaki and A. E. Kamal, "Routing techniques in wireless sensor networks: a survey," *IEEE wireless communications*, vol. 11, no. 6, pp. 6–28, 2004.
[4] J. P. Hespanha, P. Naghshtabrizi, and Y. Xu, "A survey of recent results in networked control systems," *Proceedings of the IEEE*, vol. 95, no. 1, pp. 138–162, 2007.
[5] B. Sinopoli, L. Schenato, M. Franceschetti, K. Poolla, M. I. Jordan, and S. S. Sastry, "Kalman filtering with intermittent observations," *IEEE transactions on Automatic Control*, vol. 49, no. 9, pp. 1453–1464, 2004.
[6] R. B. Myerson, *Game Theory*. Harvard university press, 2013.
[7] D. Fudenberg and J. Tirole, "Game theory (mit press)," *Cambridge, MA*, p. 86, 1991.
[8] W. Saad, Z. Han, M. Debbah, A. Hjorungnes, and T. Basar, "Coalitional game theory for communication networks," *IEEE Signal Processing Magazine*, vol. 26, no. 5, pp. 77–97, 2009.
[9] Y. Li, D. E. Quevedo, S. Dey, and L. Shi, "Sinr-based dos attack on remote state estimation: A game-theoretic approach," *IEEE Transactions on Control of Network Systems*, 2016.
[10] H. Yaïche, R. R. Mazumdar, and C. Rosenberg, "A game theoretic framework for bandwidth allocation and pricing in broadband networks," *IEEE/ACM Transactions On Networking*, vol. 8, no. 5, pp. 667–678, 2000.
[11] C. U. Saraydar, N. B. Mandayam, and D. J. Goodman, "Pricing and power control in a multicell wireless data network," *IEEE Journal on selected areas in communications*, vol. 19, no. 10, pp. 1883–1892, 2001.
[12] C. U. Saraydar, N. B. Mandayam, and D. J. Goodman, "Efficient power control via pricing in wireless data networks," *IEEE transactions on Communications*, vol. 50, no. 2, pp. 291–303, 2002.
[13] W. Tushar, W. Saad, H. V. Poor, and D. B. Smith, "Economics of electric vehicle charging: A game theoretic approach," *IEEE Transactions on Smart Grid*, vol. 3, no. 4, pp. 1767–1778, 2012.
[14] Google, "Google drive pricing homepage." https://www.google.com/drive/pricing/. Accessed May 24, 2017.
[15] B. D. Anderson and J. B. Moore, *Optimal Filtering*. Courier Corporation, 2012.
[16] L. Shi, M. Epstein, and R. M. Murray, "Kalman filtering over a packet-dropping network: A probabilistic perspective," *IEEE Transactions on Automatic Control*, vol. 55, no. 3, pp. 594–604, 2010.
[17] M. L. Puterman, *Markov Decision Processes: Discrete Stochastic Dynamic Programming*. New York, NY, USA: John Wiley & Sons, Inc., 1st ed., 1994.